\documentclass{desyproc}
\usepackage[gen]{eurosym}
\usepackage{xspace}

\def\Offline{\mbox{$\overline{\textrm%
{Off}}$\hspace{.05em}\protect\raisebox{.4ex}%
{$\protect\underline{\textrm{line}}$}}\xspace}

\begin{document}

\title{Prospects of GPGPU in the Auger Offline Software Framework}

\author{{\slshape 
	Tobias Winchen$^1$, Marvin Gottowik$^1$, and Julian Rautenberg$^1$ for the Pierre Auger Collaboration$^{2,3}$}\\[1ex]
$^1$ Bergische Universit{\"a}t Wuppertal, Gau{\ss}str. 20, 42119 Wuppertal, Germany \\
$^2$ Pierre Auger Observatory, Av. San Mart\'in Norte 304, 5613 Malarg{\"u}e, Argentina.\\ $^3$ Full author list: \url{http://www.auger.org/archive/authors_2014_09.html} }

\contribID{26}

\confID{7534}  
\desyproc{DESY-PROC-2014-05}
\acronym{GPUHEP2014} 
\doi  

\maketitle

\begin{abstract}
The Pierre Auger Observatory is  the currently largest experiment dedicated to
unveil the nature and origin of the highest energetic cosmic rays.  The
software framework \Offline has been developed by the Pierre Auger
Collaboration for joint analysis of data from different detector
systems used in the observatory. While reconstruction modules are specific to
the Pierre Auger Observatory components of the \Offline framework are also used
by other experiments. The software framework has recently been extended to
analyze also data from the Auger Engineering Radio Array (AERA), the radio
extension of the Pierre Auger Observatory.  The reconstruction of the data from
such radio detectors requires the repeated evaluation of complex antenna gain
patterns which significantly increases the required computing resources in the
joint analysis.  In this contribution we explore the usability of massive
parallelization of parts of the \Offline code on the GPU. We present the result
of a systematic profiling of the joint analysis of the \Offline software
framework aiming for the identification of code areas suitable for
parallelization on GPUs.  Possible strategies and obstacles for the usage of
GPGPU in an existing experiment framework are discussed.
\end{abstract}

\section{Introduction}
Although cosmic rays have been intensively studied for more than 100~years,
fundamental questions about the phenomenon remain unclear.  In particular, the
origin, the acceleration mechanism, and the chemical composition of the highest
energetic cosmic rays with energies up to several hundred EeV
(1~EeV~=~$10^{18}$~eV) remain open questions. As the flux of the highest
energy cosmic rays is of the order of one particle per square kilometer per 
century, direct observation is impracticable. Instead, the Earth's atmosphere
is used as a calorimeter in which the cosmic rays are detected indirectly by
the particle cascades, or `air showers', they induce. For detailed reviews on
cosmic rays see e.g.~references \cite{Kotera2011, Letessier-Selvon2011}.

The currently largest detector for cosmic rays at the highest energies is
the Pierre Auger Observatory in Argentina. The Pierre Auger Observatory~\cite{PAO2004} is
designed as a hybrid of two complementary detector systems. The `surface
detector'~\cite{Allekotte2008} consists of 1660~water-Cherenkov stations that
sample the secondary particles at the ground level. The stations are arranged in a
hexagonal grid that covers an area of 3000~km$^2$. The surface detector array
is surrounded by 27~telescopes stationed at four sites to detect the
fluorescence light emitted by air molecules that have been excited by the
particle cascade. This `fluorescence detector'~\cite{Abraham2010} provides a
direct calorimetric measurement independent of hadronic interaction models and
thus a better energy resolution than the surface detector. Furthermore, as it
measures the development of the shower in the atmosphere, it is sensitive to the
mass of the primary cosmic ray. However, the fluorescence detector can operate
only during clear and moonless nights, whereas the surface detector has no
principal constraint on the uptime.

As complementary detector without principal time constraints, the Auger Engineering Radio
Array (AERA)~\cite{Neuser2014} detects the radio pulses emitted by the particle
cascades, due to geomagnetic and charge-excess effects~\cite{Aab2014a}.
Currently, AERA consists of 124 stations in a hexagonal grid covering an area
of about 6 km$^2$. Each station is equipped with an antenna designed for
polarized measurements from 30 to 80 MHz. 

The Auger \Offline Software Framework~\cite{Argiro2007} provides the tools and
infrastructure for the reconstruction of events detected with the Pierre Auger
Observatory. Components of the framework are also used by other experiments~\cite{Sipos2012}. It is designed to support and incorporate the ideas of physicists
for the projected lifetime of the experiment of more than 20 years. This is
achieved by a strict separation of configuration, reconstruction modules, event
and detector data, and utilities to be used in the algorithms.  The
reconstruction of radio events detected with AERA is fully integrated in the
Auger \Offline Software Framework which allows joint analyses of events from all
detector systems~\cite{PAO2011g}. The main reconstruction algorithm for the
radio reconstruction and its performance profile is described in the next
section.

\section{Radio Reconstruction in the Auger Offline Framework}
\begin{figure}[tb]
	\includegraphics[width=\textwidth]{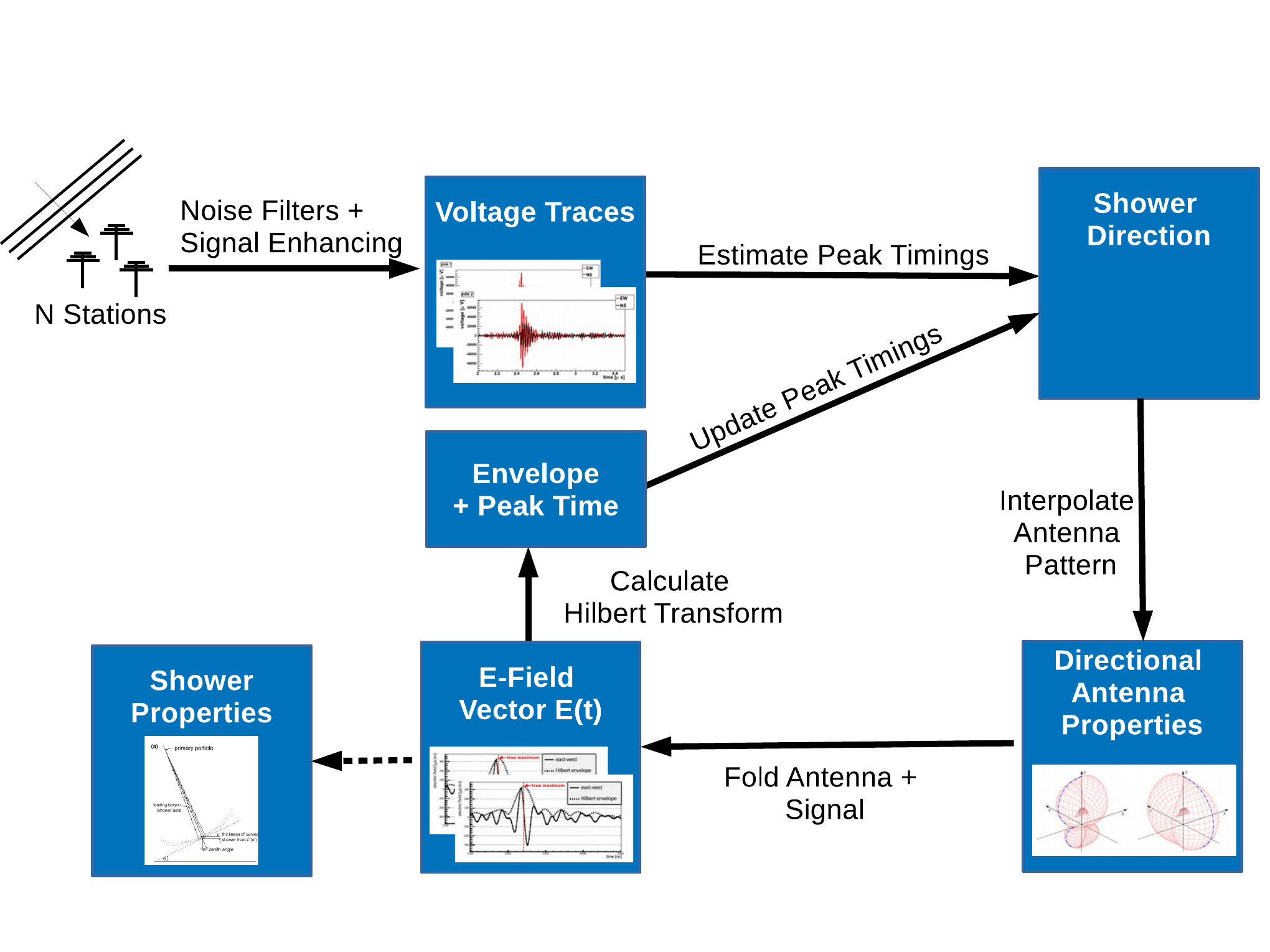}
	\caption{Schematic of the individual steps in the \Offline radio reconstruction.}
	\label{fig:OfflineRadioReconstruction}
\end{figure}

The main algorithm for the reconstruction of radio events implemented in
several \Offline modules is depicted in
Fig.~\ref{fig:OfflineRadioReconstruction}. After a trigger, the voltage traces
 of the two channels at each station are read out, and,  after noise
filtering and signal enhancing, processed as follows. First, as an initial estimate of the
event timing in each station, the maxima of the voltage traces is used. Second,
from the timing information of the individual stations, the direction of the
shower is obtained by triangulation. Third, the antenna pattern for this
direction is evaluated, and the E-field trace at each station reconstructed
from the voltage traces. Finally, the maximum of the envelope of the E-field trace is
used as updated timing information for a new iteration of the procedure. On
convergence, the timing information yields the incident direction of the
primary particle whereas the E-field distribution on the ground allows
a derivation of the energy and particle type of the cosmic ray. \\

The execution of this algorithm in \Offline requires an uncomfortable amount of
time.  Using the Linux kernel profiler `perf'~\cite{linux_perf} we identified
two main performance bottlenecks in this algorithm. First, about 15\% of the
computation time is spent in calculating Fourier transformations with the
FFTW library~\cite{fftw}. Second, about 25\% of the time is used for the
interpolation of the antenna patterns. All other parts of the algorithm use
less than 5\% of the time.  The same bottlenecks are identified using
`google-perftools'~\cite{google_perf} or `Intel VTune
amplifier'~\cite{intel_vtune} as alternative profilers.

In the next sections we discuss the elimination of these bottlenecks by
implementing the relevant parts on the GPU using the Cuda framework. In both cases
we followed  a minimum invasive approach that leaves the general interfaces in
\Offline intact. To select between the CPU and GPU implementation, a
preprocessor directive is used.

\section{Interpolations of Antenna Patterns}
To reconstruct the electric field vector from the measured voltage traces the
antenna pattern in the direction of the electromagnetic wave must be known. The antenna pattern can
be conveniently expressed as a two dimensional complex vector, the `vector
effective length (VEL)'. For each antenna, the VEL is known for discrete
frequencies and zenith and azimuth angles from measurements or simulations.
Between theses nodes the VEL has to be interpolated for arbitrary directions.

The interpolation of textures is a core task of graphics cards, which 
have dedicated circuits for the interpolation. The usage of these promises  a
great speedup, but is available for single precision data of limited size only.
In the baseline implementation, the antenna pattern is evaluated in double
precision. As a linear interpolation requires six elementary operations, the
maximum relative uncertainty from the limited floating-point precision can be
estimated as $2.5{\times}10^{-4}$\,\% in single precision. This is smaller than
other uncertainties in the reconstruction and thus negligible here.  The
largest antenna pattern considered here consists of complex data for the vector
effective length at 98 frequencies, 31 zenith angles, and 49 azimuth angles. In
single precision this corresponds to approximately 2.4~MB of data which is small
compared to the total available memory size of a few GB on modern GPUs and much
below the maximal size of a 3D texture of typically at least
2048$\times$2048$\times$2048 elements. The patterns for all antennas used in
AERA can be stored simultaneously on the device.

To speed up the interpolation on the CPU, the pattern has already been buffered in the
baseline implementation for look up on repeated access. In the GPU implementation this
is unnecessary. Here, the patterns are copied and converted to allow binding to
texture memory only once on first access.

\section{Fourier Transformations and Hilbert Envelopes}
In the baseline implementation in \Offline, calls to the FFTW library are
wrapped in an object-oriented interface. The interface provides several
distinct classes for the operation on real or complex data of given
dimensionality. Shared functionality is implemented in a base class and
propagated by inheritance. In the radio reconstruction, the wrapper operates
within a container that stores a trace simultaneously in the time and frequency
domains.  After modification of either, the other is updated in a lazy
evaluation scheme. 

To calculate the FFT on the GPU, FFTW calls are replaced by calls to the CUDA
FFT library (CuFFT). In contrast to the CPU implementation, all instances of
the GPU implementation share static memory on the GPU to avoid time consuming
allocations.  This is safe here as memory copies are blocking and performed
immediately before and after the FFT calculation. 
\begin{figure}[tb]
	\includegraphics[width=\textwidth]{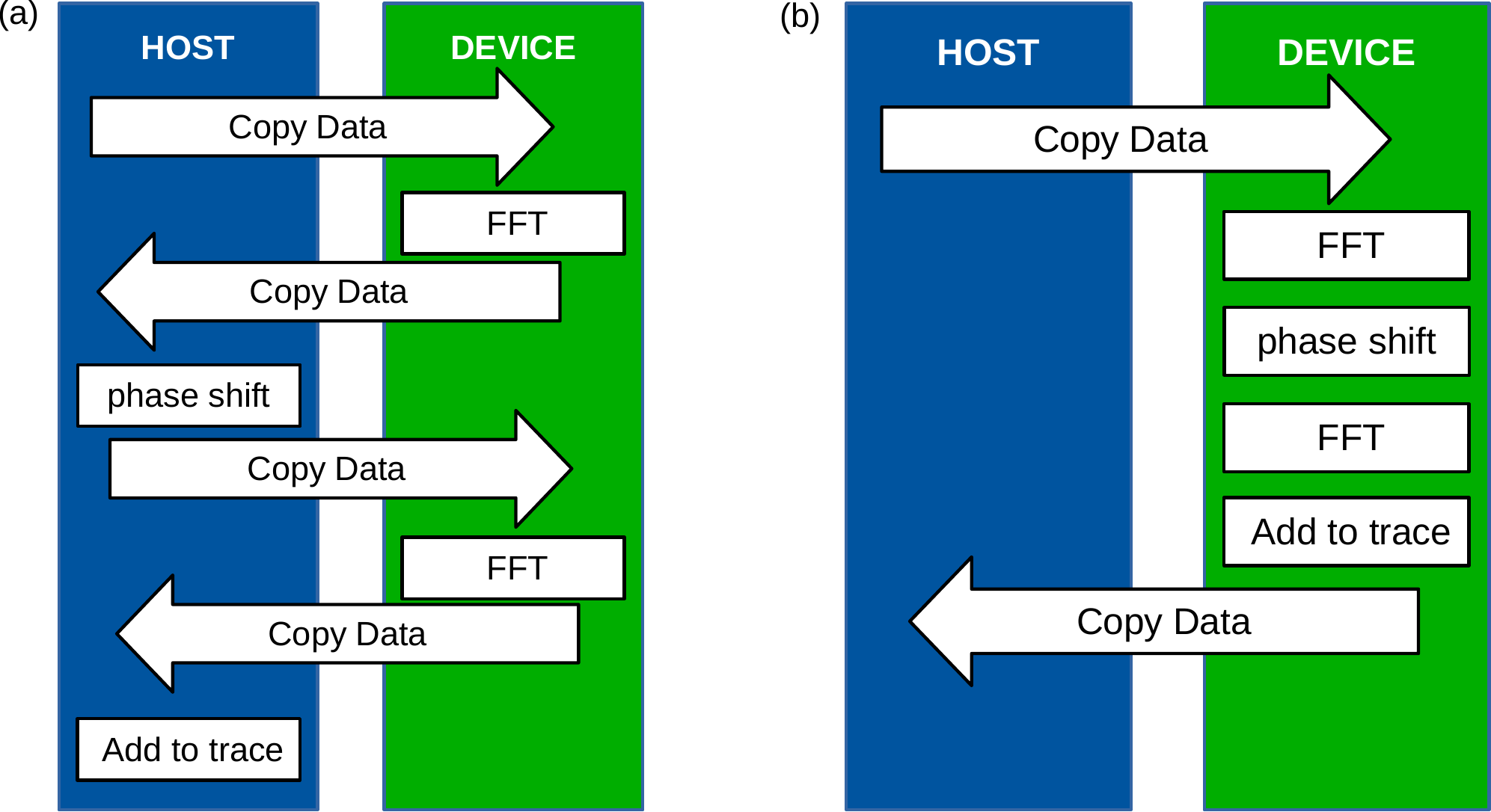}
	\caption{Calculation of the Hilbert envelope~\textbf{(a)} using the CuFFT wrapper only and \textbf{(b)} using a dedicated kernel.}
	\label{fig:HilbertImplementations}
\end{figure}

However, in the reconstruction several FFTs are calculated in context of obtaining
the Hilbert envelope of the radio traces. The envelope $E(t)$ of a trace $x(t)$
is $E(t) = \sqrt{x(t)^2 + H(x(t)^2)}$ with the Hilbert transform of the
signal $H(x(t))$. The Hilbert transformation shifts the phase of negative
frequencies, i.e.\ for band-limited signal frequencies below the
mid-frequency, by $-90^\circ$ and positive frequencies by $+90^\circ$.  The GPU
implementation as described above thus results in a non-optimal memory access
pattern for the envelope calculation as shown in
Fig.~\ref{fig:HilbertImplementations}~(a). However, with a dedicated
computing kernel not only can two memory copies be avoided, but also the phase
shift and summation are calculated in parallel
(cf.~Fig~\ref{fig:HilbertImplementations}~(b)).

\section{Discussion}
\begin{figure}[tb]
	\includegraphics[width=\textwidth]{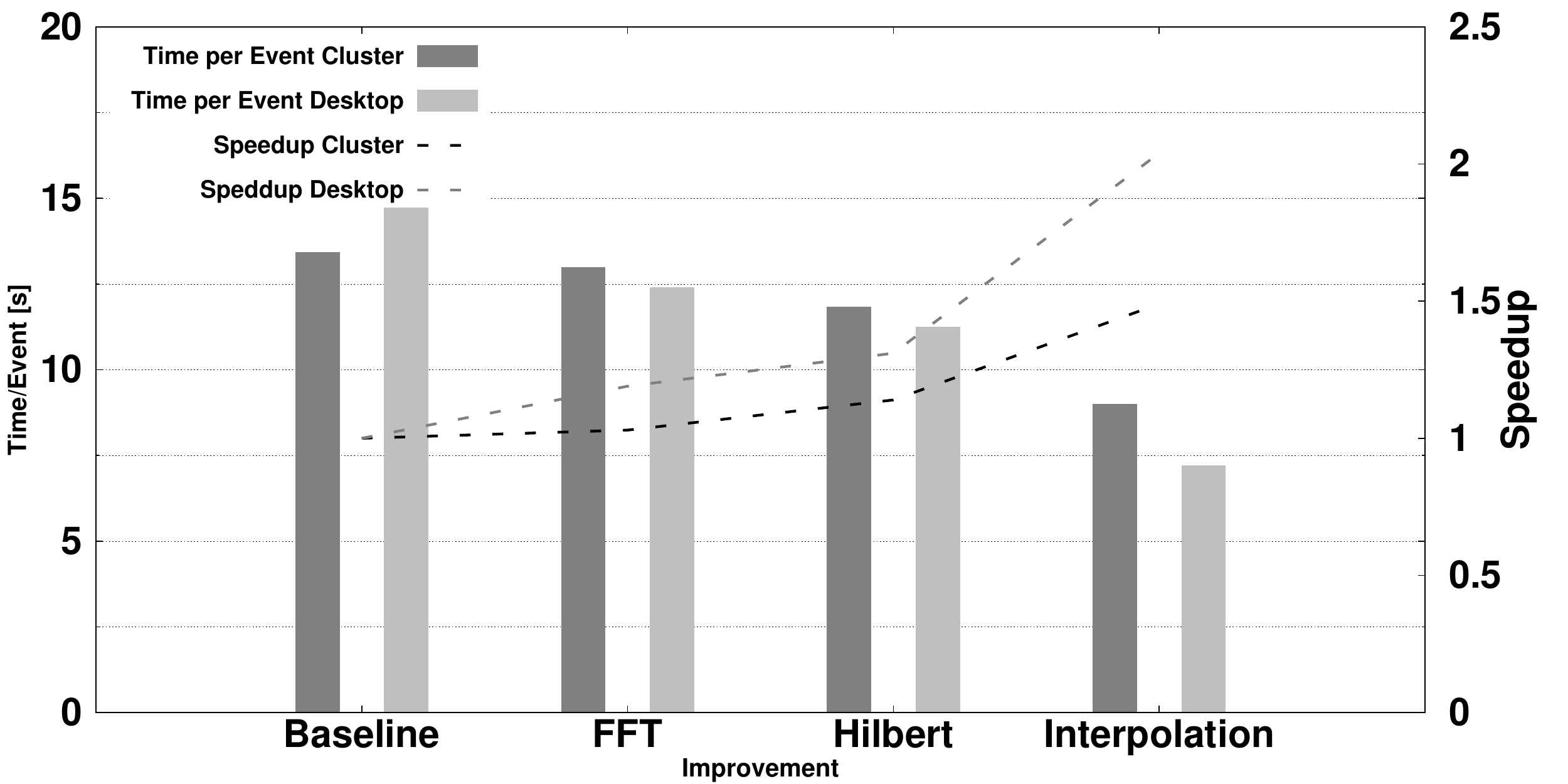}
	\caption{Summary of the speedup achieved in the individual improvements.}
	\label{fig:SpeedupSummary}
\end{figure}
The results obtained from the new GPU implementations are consistent with the results from the baseline implementation.  While the FFTs
yield identical results on CPU and GPU, the interpolation is not only a
different implementation but also only in single precision. This amounts to a
relative difference in the directional antenna patterns between the individual implementations of typically below $\pm
0.8\%$ and thus small compared to other uncertainties. 

The performance improvements obtained by the modifications are summarized in
Fig.~\ref{fig:SpeedupSummary}. As test systems we used here a typical recent
desktop PC and a combined CPU/GPU cluster.  The desktop is equipped with an AMD
A8-6600K processor and an NVIDIA GeForce 750 Ti graphics card. The cluster
contains 4 Intel Xeon X5650 CPUs and 4 NVIDIA Tesla M2090 GPUs\@.  On the desktop
system the GPU implementation of FFT and the Hilbert transformation yield a
speedup of 1.3, doing also the interpolations on the GPU increased this speedup
to approximately 2. On the cluster system the total achieved speedup is 1.5.
The lower speedup on the cluster system is due to the higher relative performance of the cluster CPU and GPU compared to the desktop system.

As only selected isolated parts of the code are moved to the GPU, the time used
for computing on the GPU is low compared to the time needed for memory copy,
and also only 7\% of the copy-time is overlapped by computing time. However,
increasing the GPU utilization would require the traces to be kept permanently
on the GPU so that more analysis steps can benefit from porting to the GPU\@.
This, however, would require non-trivial changes in the \Offline framework, in
particular, modifications of the internal structure and interfaces.

\section{Conclusion} 
The calculation of Fourier transformations and the interpolation of antenna
response patterns have been identified as bottlenecks in the AERA event
reconstruction using a performance profiler.  Eliminating both by re-implementating
the calculation in CUDA while keeping the structure of \Offline intact yields
a speedup of 1.49 to 2.04 depending on the test system.  The largest speedup is
obtained here on a typical desktop PC equipped with an entry level graphics
card. Considering the relative costs of about \euro{500} for a desktop PC and
\euro{100} for an entry level GPU, even such selected applications of GPGPU in
existing frameworks are a possibility to be considered in planning future
computing strategies.


\begin{footnotesize}

\bibliographystyle{unsrt}
\bibliography{winchen_tobias.bib}

\begin{thebibliography}{10}

\bibitem{Kotera2011}
K.~Kotera and A.~V. Olinto.
\newblock The astrophysics of ultrahigh energy cosmic rays.
\newblock {\em Annual Review of Astronomy and Astrophysics}, 49:119--153, 2011.

\bibitem{Letessier-Selvon2011}
A.~Letessier-Selvon and T.~Stanev.
\newblock {Ultrahigh Energy Cosmic Rays}.
\newblock {\em Reviews of Modern Physics}, 83:907--942, 2011.

\bibitem{PAO2004}
{J. Abraham et al. (The Pierre Auger Collaboration)}.
\newblock Properties and performance of the prototype instrument for the
  {Pierre Auger Observatory}.
\newblock {\em Nuclear Instruments and Methods in Physics Research Section A},
  523:50, 2004.

\bibitem{Allekotte2008}
{I. Allekotte et al. (The Pierre Auger Collaboration)}.
\newblock {The Surface Detector System of the Pierre Auger Observatory}.
\newblock {\em Nuclear Instruments and Methods in Physics Research Section A},
  586:409--420, 2008.

\bibitem{Abraham2010}
{J. Abraham et al. (The Pierre Auger Collaboration)}.
\newblock {The Fluorescence Detector of the Pierre Auger Observatory}.
\newblock {\em Nuclear Instruments and Methods in Physics Research Section A},
  620:227--251, 2010.

\bibitem{Neuser2014}
{J. Neuser for the Pierre Auger Collaboration}.
\newblock {Detection of Radio Emission from Air Showers in the MHz Range at the
  Pierre Auger Observatory}.
\newblock In {\em Proceedings of the 6th international Conference on Acoustic
  and Radio EeV Neutrino Detection Activities (ARENA)}, 2014.

\bibitem{Aab2014a}
{A. Aab et al. (The Pierre Auger Collaboration)}.
\newblock Probing the radio emission from air showers with polarization
  measurements.
\newblock {\em Phys. Rev. D}, 89:052002, Mar 2014.

\bibitem{Argiro2007}
{S. Argiro et al.}
\newblock {The Offline Software Framework of the Pierre Auger Observatory}.
\newblock {\em Nuclear Instruments and Methods in Physics Research Section A},
  580:1485--1496, 2007.

\bibitem{Sipos2012}
{R. Sipos et al.}
\newblock The offline software framework of the na61/shine experiment.
\newblock In {\em Proceedigs of the International Conference on Computing in
  High Energy and Nuclear Physics 2012 (CHEP2012)}, 2012.

\bibitem{PAO2011g}
{P. Abreu et al. (The Pierre Auger Collaboration)}.
\newblock {Advanced functionality for radio analysis in the Offline software
  framework of the Pierre Auger Observatory}.
\newblock {\em Nuclear Instruments and Methods in Physics Research Section A},
  A635:92--102, 2011.

\bibitem{linux_perf}
{The Linux Kernel Team}.
\newblock Perf wiki.
\newblock \url{https://perf.wiki.kernel.org/index.php/Main_Page}, visited on
  2014-09-19.

\bibitem{fftw}
M.~Frigo and S.~G. Johnson.
\newblock The design and implementation of {FFTW3}.
\newblock {\em Proceedings of the IEEE}, 93(2):216--231, 2005.
\newblock Special issue on ``Program Generation, Optimization, and Platform
  Adaptation''.

\bibitem{google_perf}
{Google Inc.}
\newblock gperftools - fast, multi-threaded malloc() and nifty performance
  analysis tools.
\newblock \url{https://code.google.com/p/gperftools/}, visited on 2014-08-10.

\bibitem{intel_vtune}
{The Intel Corporation}.
\newblock {Intel VTune Amplifier}.
\newblock \url{https://software.intel.com/en-us/intel-vtune-amplifier-xe},
  visited on 2014-08-02.

\end{thebibliography}


\end{footnotesize}
\end{document}